# Three-dimensional solid-state qubit arrays with long-lived spin coherence

C.J. Stephen[1], B.L. Green[1], Y.N.D. Lekhai[1], L. Weng[2], P. Hill[3,4], S. Johnson[2], A.C. Frangeskou[1], P.L. Diggle[1,4], M.J. Strain[3,4], E. Gu[3,4], M.E. Newton[1,4], J.M. Smith[2,4], P.S. Salter[5] & G.W. Morley[1,4]

[1]Department of Physics, University of Warwick, Gibbet Hill Road, Coventry, CV4 7AL, UK.

[2]Department of Materials, University of Oxford, Parks Road, Oxford OX1 3PH, UK.

[3]Institute of Photonics, University of Strathclyde, George Street, Glasgow, G1 1RD, UK.

[4]EPSRC Centre for Doctoral Training in Diamond Science and Technology, University of Warwick, Coventry, CV4 7AL, UK.

[5]Department of Engineering Science, University of Oxford, Parks Road, Oxford OX1 3PJ, UK.

Three-dimensional arrays of silicon transistors increase the density of bits[1]. Solid-state qubits are much larger so could benefit even more from using the third dimension given that useful fault-tolerant quantum computing will require at least 100,000 physical qubits and perhaps one billion[2]. Here we use laser writing to create 3D arrays of nitrogen-vacancy centre (NVC) qubits in diamond. This would allow 5 million qubits inside a commercially available 4.5x4.5x0.5 mm diamond based on five nuclear qubits per NVC[3,4] and allowing (10 µm)$^3$ per NVC to leave room for our laser-written electrical control. The spin coherence times we measure are an order of magnitude longer than previous laser-written qubits[5] and at least as long as non-laser-written NVC[6]. As well as NVC quantum computing[3,4,6-8], quantum communication[7,9,10] and nanoscale sensing[11-14] could benefit from the same platform. Our approach could also be extended to other qubits in diamond[15-18] and silicon carbide[19,20].

Demonstrated qubit fidelities[8] for a single negatively-charged nitrogen vacancy centre (NVC) and its nearby nuclear spins are above the required thresholds for quantum computing[2]. Two NVCs in different diamonds, in separate cryostats, have been optically entangled faster than the decoherence of this entanglement[7], but it will not be practical to have $10^6$ cryostats for $10^6$ NVCs. In the transparent lattice of wide-band-gap diamond, individual optically-addressable qubits can fill a volume rather than be restricted to the surface. For computation, a 3D array spanning the upper 50 µm of a commercially-available electronic (EL) grade 4.5×4.5×0.5 mm diamond could contain $10^6$ NVCs with (10 µm)$^3$ for each NVC. Each NVC has, on average, five individually-addressable $^{13}$C nuclear spin qubits[3,4]. For communications, having an array of NVCs will provide many spin-photon interfaces within one cryostat[10], increasing data rates and allowing multiplexing. Sensing with 2D arrays of NVCs will combine the high resolution of single NVC sensing[11] with the simultaneous imaging achieved with wide-field microscopy[13]. Stacking two of these 2D arrays will then permit gradiometry which will increase the sensitivity by subtracting the background noise measured by the array that is further from the sample of interest.

For all these technologies, we envision a fibre bundle or a spatial light modulator (SLM) that sends and receives optical photons to and from a 2D array of NVCs simultaneously, through a common lens. Moving the diamond closer to the lens would allow 2D arrays at different depths to be probed sequentially. A bullseye grating array would be used to collect more of the fluorescence[21]. Time would not be wasted even for 50 2D layers, as optical initialisation/readout of each NVC only lasts 3 µs, followed by a delay that is on the order of the NVC electron spin coherence time, $T_2$, which should be at least 500 µs at room temperature and longer in a cryostat. For sensing, delta-doping with nitrogen would be used to control the depth of NVCs with a precision of 4 nm[22].

The creation of 2D NVC arrays has been demonstrated previously[5,22-24]. With ion implantation through a mask, high-precision placement of 10 nm[23] has been shown, and electron spin $T_2$ times of up to 50 µs. Longer $T_2$ times of up to 530 µs were achieved by using isotopically pure $^{12}C$ diamond but with less precise placement and again requiring ion implantation through a mask[24]. For quantum computing it is important to use diamond with natural isotopic abundance because the 1.1% $^{13}C$ nuclear spins provide a valuable register of around five qubits that can be used to store quantum information for longer than the electron spin. Localised electron irradiation into a diamond provided a 2D array with $T_2$ of up to 1.3 ms due to the use of $^{12}C$ diamond, and no need for a mask[22]. This used delta doping to reach a depth precision of 4 nm, with in-plane precision of 450 nm. It has been shown that 2D arrays of NVCs can be laser written with no mask, but the $T_2$ time measured was typically only 30 to 80 µs[5]. A new preprint demonstrates preferential orientation and near-100% yield for 5x5 2D arrays of laser-written NVC in diamond[25].

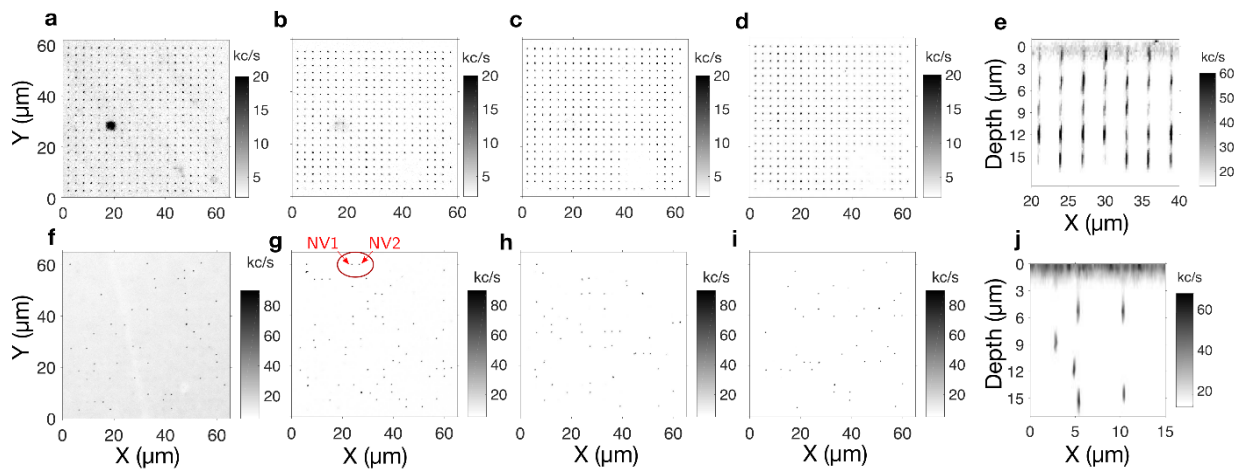

Fig. 1. Confocal imaging of a three dimensional array of defects in diamond (array M). Top row: **a**, **b**, **c** and **d** are in the XY plane at depths of 6, 9, 12 and 15 µm respectively before annealing: each spot is an ensemble of vacancies. **e** is a vertical section in the XZ plane. Bottom row: The same volume after annealing out the vacancies: the spots are nitrogen vacancy centres (NVCs). In 9% of the target sites NVCs were created, almost all as single NVCs.

Our 3D arrays were created by laser-writing over 2000 NVCs into a diamond with natural isotopic abundance of $^{13}$C as shown in Fig. 1. We measured the spin echo coherence time for 23 of the single NVC and found that 16 of them had $T_2$ >500 µs at room temperature. All our measurements are at room temperature where the electron spin coherence time is limited by the natural isotopic abundance of $^{13}$C. We used dynamic decoupling to probe the electron spin coherence without $^{13}$C limitations, finding a coherence time limited by the electron spin $T_1$ as has been reported for naturally occurring NVCs.

To create the arrays, an EL grade diamond was bought from Element 6, and plasma etching was used to remove 20 µm of sub-surface polishing damage[14]. Arrays of ensembles of vacancies were generated in the diamond lattice by single 250 fs pulses from a 790 nm laser focused tightly beneath the surface of the diamond using a high numerical aperture (NA) oil objective. The light matter interaction is highly non-linear, limiting any material modification to the centre of the focal volume and giving an inherent three-dimensional resolution to the fabrication[26]. The refractive index mismatch at the oil-diamond interface causes refraction leading to a depth dependent spherical aberration of the laser focus, which can limit three-dimensional fabrication resolution. Adaptive optics using a liquid crystal SLM were used to correct for the aberration[27], ensuring that the fabrication was the same at each depth. The full-width-half-maxima of the intensity distribution of the laser focus inside the diamond are theoretically estimated to be 350 nm radially and 1.7 µm longitudinally. However, the expected dimensions over which the light matter interaction is appreciable are likely to be much lower[28]. The diamond sample was mounted on a three axis precision translation stage and moved relative to the laser focus to fabricate arrays of vacancy ensembles.

In order to find the fine range of pulse energies to use in these experiments, an initial calibration study was carried out on the same fabrication run inside a nominally identical diamond by writing arrays of points across a coarse range of pulse energies. This sample was subsequently characterised using a scanning confocal microscope, to find the pulse energy that produced just-visible vacancy ensembles using an air objective. Previous work has shown that write-pulse energies slightly lower than this are optimal for NVC creation[5]. Twenty-one 3D arrays labelled A to V were laser written with different energies from 14 to 19 nJ and with different pitches from 2 to 5 µm, with each 3D array having 21×20 2D arrays stacked with up to five depths for a total of over 44,000 writing sites.

Figure 1 contains images from our scanning confocal microscope of one of the 3D arrays (array M) before and after annealing. Before annealing, laser-written spots are visible due to the fluorescence of neutrally charged vacancies $V^0$: lattice sites in the diamond with missing carbon atoms. This is shown in the top row of Fig. 1. The characteristic $V^0$ fluorescence spectrum confirms the identity of these vacancies as shown in the Supplementary Information. None of these pre-anneal spots were visible with our air objective[5] but these images were collected with our oil objective (NA=1.4). We annealed the diamond at 1000˚C for 3 hours in a nitrogen environment[5] and repeated the imaging, as shown in the bottom row of Fig. 1. The spots that can be seen are NVCs as confirmed by the characteristic fluorescence spectrum shown in the Supplementary Information. We have analysed the precision of the position for 167 single NVCs by three-dimensional fitting of the point spread function as shown in Fig. 2. This reveals that the NVCs are in the desired locations to within ±200 nm in the transverse (XY) plane and ±250 nm in the vertical (Z) direction. The high precision in the Z direction is due to the non-linearity of the writing. The precision is probably limited by the concentration of nitrogen in this material which is in the range from 1 to 5 ppb corresponding to an average spacing between nitrogen atoms of 180 to 100 nm. This implies that the vacancies generally bond to one of their nearest nitrogen dopants.

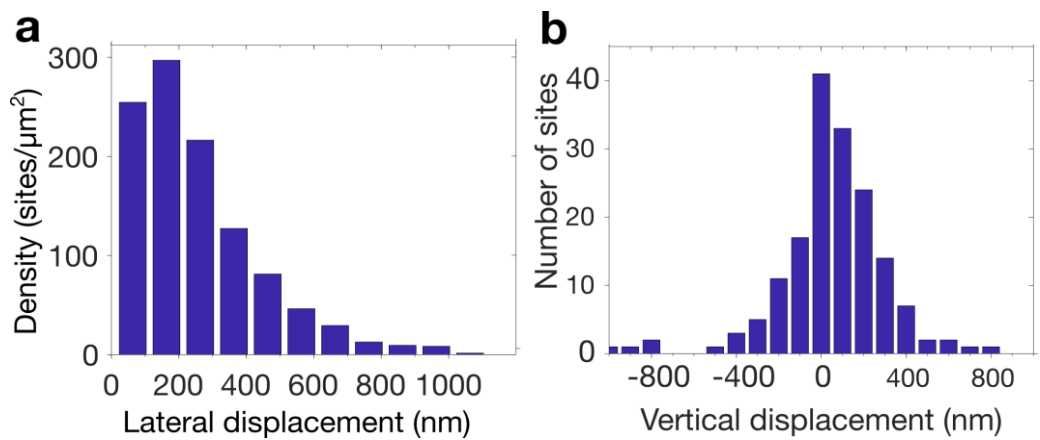

Fig. 2. Positioning precision of single nitrogen vacancy centres in array M. **a** in the XY plane of the array. **b** in the vertical Z direction.

To check if the NVCs are single centres we used automated Hanbury-Brown Twiss (HBT) experiments on over 600 sites to measure the photon arrival autocorrelation function $g^2(\tau)$ as shown in Fig. 3. We classify a site as a single emitter where $g^2(0) < 0.5$, a double for $0.5 \leq g^2(0) < 0.66$ and a triple for $0.66 \leq g^2(0) < 0.75$. For array M, 87% of the NVCs are single centres, with 11% being doubles and 2% triples. The sites with no NVCs could be repeatedly re-written with another laser pulse and re-annealed until no sites are empty. The other paper to report single NVC creation with laser writing [5] used higher pulse energies and reported a higher yield of single NVCs but there were similarly more of the unwanted double and triple NVCs which would lead to a lower yield of single NVCs in a repeat-until-success strategy. The Supplementary Information contains confocal imaging and HBT statistics for some of the other arrays. Of the 2050 sites written in array M, 8% developed into a single NVC, almost 1% produced a double NVC and <0.1% produced a triple NVC. This is consistent with Poissonian statistics (as $0.01 \approx 0.08^2$ and $0.001 \approx 0.08^3$) suggesting that NVC creation is limited by the nitrogen density rather than by the highly non-linear laser writing.

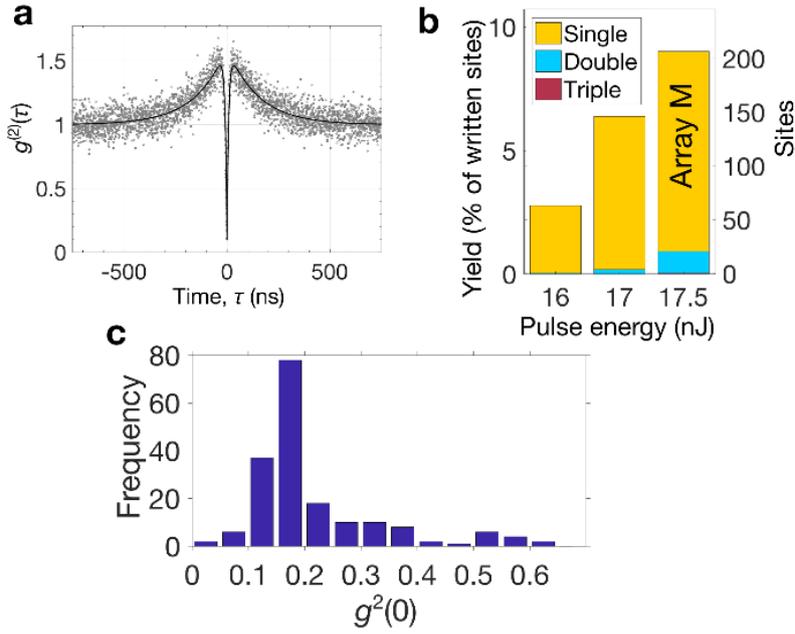

Fig. 3. Fraction of single nitrogen vacancy centres compared to doubles and triples. **a** Hanbury-Brown Twiss (HBT) measurement of the photon arrival time for NV1. **b** HBT measurements reveal the fraction of single, double and triple NVCs produced in three of the 3 μm pitch arrays with different laser-write pulse energies. **c** The frequency of the measured $g^2(0)$ for array M with no background subtraction.

Figure 4 shows measurements of the spin coherence from 23 of the single NVCs in the 3 μm pitch arrays M and I. The longest room-temperature spin-echo coherence times without $^{12}C$ enrichment we have found in the literature are $T_2$ = 687 μs[6] and $T_2$ = 650 μs[29], which are slightly below (but within the error of) our five longest times. The long times we measure demonstrate that our laser-writing technique does not introduce excess damage or impurities to the environment of the NVCs. Our calibration step to ensure we used the optimum write-pulse energy may be needed to achieve this[5].

Previous NVC optical entanglement work has applied electric fields to Stark shift the optical florescence frequency so that that the two NVCs have indistinguishable emission[6,7]. Fig. 4c shows an NVC between two electrically-conducting wires that we laser-wrote in 3D at the same time as the arrays. It is known that these laser-written wires in diamond are graphitic and that they conduct with a DC resistivity of around 0.1 Ωcm at room temperature[30].

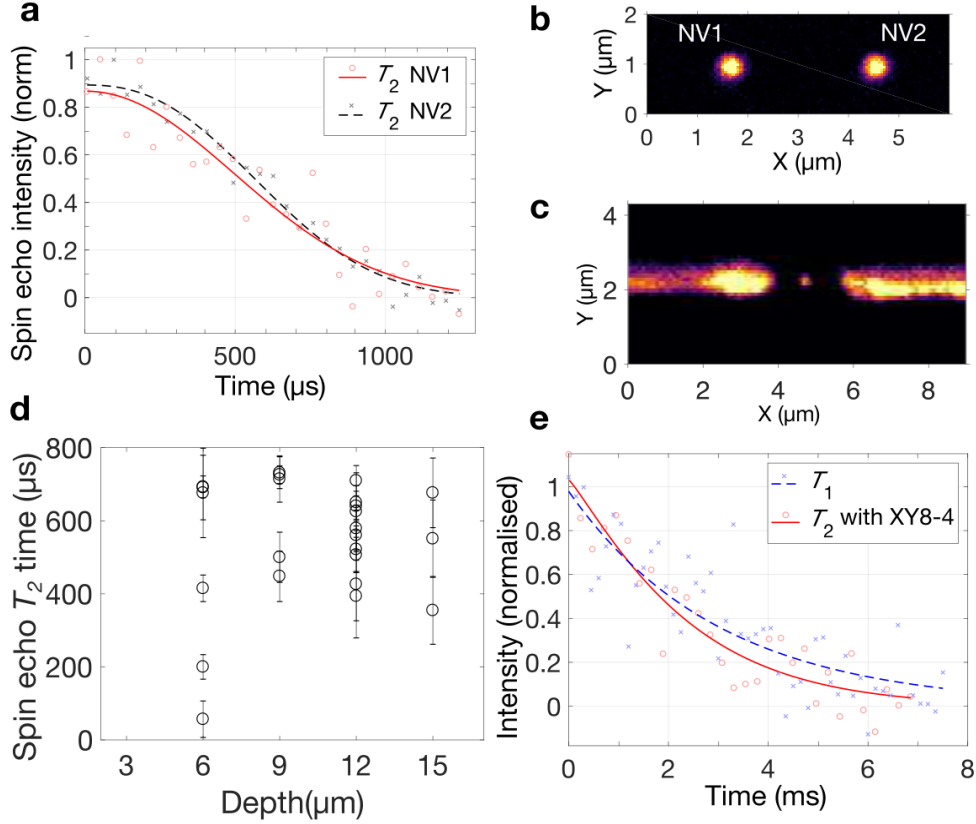

Fig. 4. Nitrogen vacancy centre electron spin coherence times. **a** Spin-echo decays for adjacent, aligned NVCs (labelled NV1 and NV2, as shown in Fig. 4b and Fig. 1g) fitted with $Ae^{-\left(\frac{t}{T_2}\right)^n}$. NV1 has $T_2$ = 690±80 μs with $n$ = 2.0±0.8, while NV2 has $T_2$ = 710±40 μs with $n$ = 2.4±0.4. **b** Zoomed confocal image of NV1 and NV2. **c** Confocal image of an NVC between two laser written electrical wires from another region of the same diamond. **d** Histogram of spin echo times measured for sites in arrays M (17.5 nJ) and I (17 nJ) which had a 3 μm pitch, as a function of NVC depth. **e** Using XY8-4 dynamic decoupling achieved $T_2$ = 2.4±0.6 ms with $n$ = 1.1±0.4 on NV2. For comparison, the longitudinal lifetime, $T_1$, of this site was 3.0±0.7 ms.

Methods: All data presented were collected with a home-built scanning confocal microscope at room temperature. 532 nm light was used to excite the sample and the 637-800 nm florescence was collected with single photon counting modules. The electron spins were coherently controlled with microwaves at 2.8-3 GHz. A 25 mT magnetic field was applied along the [111] direction for the spin coherence measurements, suppressing the periodic revivals. Each spin coherence measurement takes 2 to 12 hours, depending on the desired signal to noise. The equipment is controlled using Qudi software[31] with some of our modifications. Once set up, laser writing of a 2000 point array typically takes 5 minutes.

1  Sachid, A. B. *et al.* Monolithic 3D CMOS Using Layered Semiconductors. *Advanced Materials* **28**, 2547-2554 (2016).
2  O'Gorman, J. & Campbell, E. T. Quantum computation with realistic magic-state factories. *Physical Review A* **95**, 032338 (2017).
3  Reiserer, A. *et al.* Robust Quantum-Network Memory Using Decoherence-Protected Subspaces of Nuclear Spins. *Phys. Rev. X* **6**, 021040 (2016).

4	Abobeih, M. H. *et al.* One-second coherence for a single electron spin coupled to a multi-qubit nuclear-spin environment. *Nature Communications* **9**, 2552 (2018).
5	Chen, Y.-C. *et al.* Laser writing of coherent colour centres in diamond. *Nat Photon* **11**, 77-80 (2017).
6	Bernien, H. *et al.* Heralded entanglement between solid-state qubits separated by three metres. *Nature* **497**, 86-90 (2013).
7	Humphreys, P. C. *et al.* Deterministic delivery of remote entanglement on a quantum network. *Nature* **558**, 268-273 (2018).
8	Rong, X. *et al.* Experimental fault-tolerant universal quantum gates with solid-state spins under ambient conditions. *Nature Communications* **6**, 8748 (2015).
9	Yang, S. *et al.* High-fidelity transfer and storage of photon states in a single nuclear spin. *Nat Photon* **10**, 507-511 (2016).
10	Babinec, T. M. *et al.* A diamond nanowire single-photon source. *Nat. Nanotechnol.* **5**, 195 (2010).
11	Bonato, C. *et al.* Optimized quantum sensing with a single electron spin using real-time adaptive measurements. *Nat. Nanotechnol.* **11**, 247 (2015).
12	Gross, I. *et al.* Real-space imaging of non-collinear antiferromagnetic order with a single-spin magnetometer. *Nature* **549**, 252 (2017).
13	Le Sage, D. *et al.* Optical magnetic imaging of living cells. *Nature* **496**, 486-U105 (2013).
14	Appel, P. *et al.* Fabrication of all diamond scanning probes for nanoscale magnetometry. *Rev. Sci. Instrum.* **87**, 063703 (2016).
15	Green, B. L. *et al.* Neutral Silicon-Vacancy Center in Diamond: Spin Polarization and Lifetimes. *Physical Review Letters* **119**, 096402 (2017).
16	Rose, B. C. *et al.* Observation of an environmentally insensitive solid-state spin defect in diamond. *Science* **361**, 60-63 (2018).
17	Pingault, B. *et al.* All-Optical Formation of Coherent Dark States of Silicon-Vacancy Spins in Diamond. *Physical Review Letters* **113**, 263601 (2014).
18	Rogers, L. J. *et al.* All-Optical Initialization, Readout, and Coherent Preparation of Single Silicon-Vacancy Spins in Diamond. *Physical Review Letters* **113**, 263602 (2014).
19	Koehl, W. F., Buckley, B. B., Heremans, F. J., Calusine, G. & Awschalom, D. D. Room temperature coherent control of defect spin qubits in silicon carbide. *Nature* **479**, 84-88 (2011).
20	Widmann, M. *et al.* Coherent control of single spins in silicon carbide at room temperature. *Nature Materials* **14**, 164 (2014).
21	Li, L. *et al.* Efficient Photon Collection from a Nitrogen Vacancy Center in a Circular Bullseye Grating. *Nano Lett.* **15**, 1493-1497 (2015).
22	McLellan, C. A. *et al.* Patterned Formation of Highly Coherent Nitrogen-Vacancy Centers Using a Focused Electron Irradiation Technique. *Nano Lett.* **16**, 2450-2454 (2016).
23	Scarabelli, D., Trusheim, M., Gaathon, O., Englund, D. & Wind, S. J. Nanoscale Engineering of Closely-Spaced Electronic Spins in Diamond. *Nano Lett.* **16**, 4982-4990 (2016).
24	Ohno, K. *et al.* Three-dimensional localization of spins in diamond using 12C implantation. *Appl. Phys. Lett.* **105**, 052406 (2014).
25	Chen, Y.-C. *et al.* Laser writing of individual atomic defects in a crystal with near-unity yield. *arXiv* (2018).
26	Gattass, R. R. & Mazur, E. Femtosecond laser micromachining in transparent materials. *Nat. Photonics* **2**, 219 (2008).
27	Simmonds, R. D., Salter, P. S., Jesacher, A. & Booth, M. J. Three dimensional laser microfabrication in diamond using a dual adaptive optics system. *Opt. Express* **19**, 24122-24128 (2011).
28	Lagomarsino, S. *et al.* Photoionization of monocrystalline CVD diamond irradiated with ultrashort intense laser pulse. *Physical Review B* **93**, 085128 (2016).

29      Mizuochi, N. *et al.* Coherence of single spins coupled to a nuclear spin bath of varying density. *Physical Review B* **80**, 4 (2009).
30      Sun, B., Salter, P. S. & Booth, M. J. High conductivity micro-wires in diamond following arbitrary paths. *Appl. Phys. Lett.* **105**, 231105 (2014).
31      Binder, J. M. *et al.* Qudi: A modular python suite for experiment control and data processing. *SoftwareX* **6**, 85-90 (2017).

**Acknowledgements** C.J.S.'s PhD is funded by the Royal Society and G.W.M. is supported by the Royal Society. P.H.'s PhD is co-funded by the Fraunhofer Centre for Applied Photonics. P.H. and P.L.D.'s PhDs are co-funded by the EPSRC Centre for Doctoral Training in Diamond Science and Technology (EP/L015315/1). This work was supported by the EPSRC NQIT (Networked Quantum Information Technology) Hub (EP/M013243/1).